\documentclass[traditabstract, bibyear]{aa} % for the abstract without structuration 
\usepackage{graphicx}
\usepackage{txfonts}
\usepackage{color}
\usepackage{amstext}
\usepackage{answers}

\begin{document}

   \title{Signature of non-isotropic distribution of stellar rotation inclination angles in the Praesepe cluster}

   \author{Geza Kovacs\inst{1}}

   \institute{Konkoly Observatory of the Hungarian Academy of Sciences, 
              Budapest, 1121 Konkoly Thege ut. 13-15, Hungary \\
              \email{kovacs@konkoly.hu}
             }

   \date{Received 12 June 2017 / Accepted 6 March 2018}

% \abstract{}{}{}{}{} 
% 5 {} token are mandatory
 
  \abstract
%{...}
%{...}
%{...}
%{...}
{The distribution of the stellar rotation axes of $113$ main sequence 
stars in the open cluster Praesepe are examined by using current photometric 
rotation periods, spectroscopic rotation velocities, and estimated stellar 
radii. Three different samples of stellar rotation data on spotted stars from 
the Galactic field and two independent samples of planetary hosts are used as 
control samples to support the consistency of the analysis. Considering the 
high completeness of the Praesepe sample and the behavior of the control 
samples, we find that the main sequence $F - K$ stars in this cluster are 
susceptible to rotational axis alignment. Using a cone model, the most likely 
inclination angle is $76^\circ \pm 14^\circ$ with a half opening angle of 
$47^\circ \pm 24^\circ$. Non-isotropic distribution of the inclination angles 
is preferred over the isotropic distribution, except if the rotation velocities 
used in this work are systematically overestimated. We found no indication of 
this being the case on the basis of the currently available data.}

   
\titlerunning{Non-isotropic distribution of inclination angles}
\authorrunning{Kovacs, G.}
   \maketitle
%
%________________________________________________________________

%
%%%%%%%%%%%%%%%%%%%%%%%
% SECTION 1
%%%%%%%%%%%%%%%%%%%%%%%
%
\section{Introduction}
Current space- and ground-based surveys have made a significant contribution 
 to our understanding of stellar rotation (e.g., McQuillan, 
Mazeh \& Aigrain~\cite{mcquillan2014}; Delorme et al.~\cite{delorme2011}; 
Hartman et al.~\cite{hartman2010}). Because stellar rotation carries 
information  on the angular momentum evolution of the stars themselves 
and  on the whole cluster (but in a rather involved way), these 
data are also relevant  for understanding the poorly known early phase of 
cluster evolution, including the loss of angular momentum, the importance of gas 
dynamics, turbulence, and the magnetic field 
(e.g., McKee \& Ostriker~\cite{mckee2007}). 

Inspired by the recent finding of Corsaro et al.~(\cite{corsaro2017}) 
 on the apparent large degree of rotational axis alignment in two 
old clusters, here we examine the distribution of the line-of-sight 
component of the orientation of the rotational axes of $113$ main 
sequence stars in the $\sim 0.8$~Gyr old cluster Praesepe. To derive 
the inclination angles we use spectroscopic and photometric rotational 
data combined with stellar radii from stellar evolution models. 
The distribution of the  inclination angles derived in this way are modeled by 
various assumptions on their underlying distributions. 

%  
%%%%%%%%%%%%%%%%%%%%%%%
% SECTION 2
%%%%%%%%%%%%%%%%%%%%%%%
%
\section{Stellar rotation data}

\subsection{Praesepe}
The dataset we used is based primarily on the availability of spectroscopically 
derived rotational velocities. The $152$ stars listed in 
Mermilliod, Mayor \& Udry~(\cite{mermilliod2009}) with possible cluster 
membership have been cross-correlated with the $180$ stars from the 
HATNet\footnote{https://hatnet.org/} survey (Kovacs et al.~\cite{kovacs2014}) 
and with the $941$ stars recently published by Rebull et al.~(\cite{rebull2017}) 
from the Kepler two-wheel (K2) survey. After various filter steps (e.g., lack 
of ${\rm v}sini$\footnote{Throughout the paper we use the shorthand notation 
${\rm v}sini$ and $sini$, respectively, for the spectroscopic rotational 
velocity $V_{\rm eq}\sin (i)$ (with $V_{\rm eq}$ for the equatorial velocity) 
and for the sine of the inclination angle.}, stars with multiple matches) 
we ended up with $120$ stars that all have 2MASS 
(Skrutskie et al.,~\cite{skrutskie2006}) and V colors (nearly exclusively 
from APASS: Henden et al.,~\cite{henden2016}), spectroscopic and photometric 
rotational data\footnote{These data, together with the other two compiled 
datasets used in this paper, are accessible at the Strasbourg astronomical 
Data Center: \url{http://cdsweb.u-strasbg.fr/}}. We note that this sample 
is unbiased. No cut was made on the basis of the value of ${\rm v}sini$, 
period, and color. Furthermore, because of the high precision of the 
K2 observations, the period sample is nearly complete; the periodic signal 
detection rate is $86$\% for the $941$ stars examined in the work of 
Rebull et al.~\cite{rebull2017}. The rotation periods are also highly 
reliable, with an overall relative error of $3$\%, based on the comparison 
of the common stars in the Kovacs et al.~\cite{kovacs2014} and 
Rebull et al.~\cite{rebull2017} samples, separated by several years. 

%
%################
% Figure 1
%################
%
\begin{figure}
 \vspace{0pt}
 \centering
 \includegraphics[angle=-90,width=85mm]{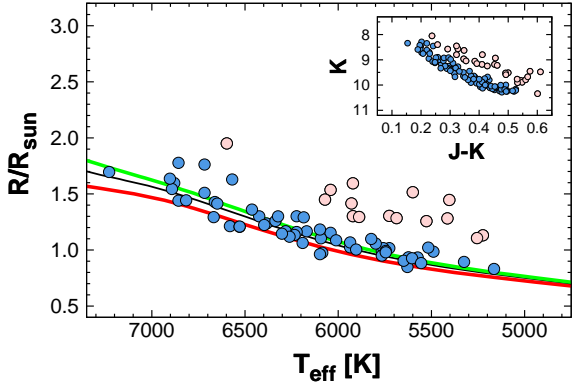}
 \caption{Comparison of the radii derived from interferometrically calibrated 
 photometric data by Bourg\'es et al.~(\cite{bourges2017}) with the $0.8$~Gyr 
 PARSEC isochrones for [Fe/H]=$0.0$, $+0.12$, and $+0.19$, plotted with red, 
 black, and green lines, respectively. The outliers are indicated by the 
 pale dots and constitute a subset of the CMD outliers shown in the 
 inset.}
\label{ldd}
\end{figure}

%
%################
% Figure 2
%################
%
\begin{figure}
 \vspace{0pt}
 \centering
 \includegraphics[angle=-90,width=85mm]{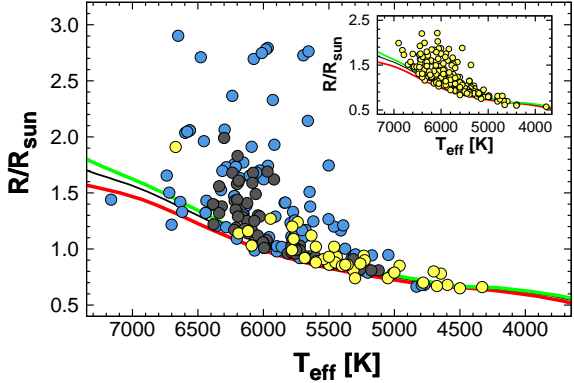}
 \caption{Comparison of the radii based on the PARSEC stellar 
 evolutionary models for Praesepe with other stellar radii derived 
 for Galactic field stars. The isochrones shown in Fig.~\ref{ldd} 
 are used. The interferometric radii collected by 
 Boyajian et al.~(\cite{boyajian2013}) are shown by blue dots, those 
 derived from the extrasolar planetary systems in the Kepler field 
 (Winn et al.~\cite{winn2017}) by deep gray dots, and hot Jupiter 
 hosts elsewhere (Maxted et al.~\cite{maxted2015}, extended/updated 
 for this study) by yellow dots. The inset shows the same kind of 
 plot, but  using the large compilation of transiting planet hosts by 
 Hartman et al.~(\cite{hartman2016}).}
\label{boyajian}
\end{figure}

%
%================
% Table 1
%================
%
\begin{table}[!h]
%\centering
  \caption{Summary of the datasets used in this paper}
  \label{datasets}
  \scalebox{0.97}{
  \begin{tabular}{lcccr}
  \hline
   Dataset       & N$_{\rm tot}$/N  & $\langle {\rm v}sini \rangle$ & $\langle\sigma({\rm v}sini)\rangle$ &  Source \\ 
 \hline
Praesepe         & 120/113          & {\phantom 0}8.4 & {\phantom 0}1.5 & {\small M09, K14, R17} \\
HJ$_{\rm host}$  &  39/39           & {\phantom 0}4.0 & {\phantom 0}0.6 & {\small M15} \\
KOI$_{\rm host}$ &  61/61           & {\phantom 0}6.2 & {\phantom 0}1.1 & {\small W17} \\
Field~(A)        &  97/91           & 51.3            & 20.0            & {\small A01} \\
Field~(B)        & 193/179          & 46.1            & 14.8            & {\small Net17} \\
Field~(C)        &  55/46           & 24.8            & {\phantom 0}1.6 & {\small M13, N15, N17} \\
                 &                  &                 &                 & {\small N13, R15, Mc14} \\
\hline
\end{tabular}}
\begin{flushleft}
{\bf Notes:}\\
$\bullet$ HJ$_{\rm host}$ and KOI$_{\rm host}$ stand for the two separate 
          sets of extrasolar planet host stars (mostly with hot Jupiters 
          and various Kepler planets, respectively)\\ 
$\bullet$ Field(A,B,C) stand for three sets of Galactic field stars\\ 
$\bullet$ $\langle {{\rm v}sini} \rangle$ and $\langle\sigma({\rm v}sini)\rangle$, 
          respectively, denote the averages of the  ${{\rm v}sini}$ values 
          and the averages of their errors given in the source for ${{\rm v}sini}$. 
          The total number of items and the number  actually 
          used in this work are denoted by N$_{\rm tot}$ and N, respectively. 
          See Sect.~2.2 for the description of the selection criteria.\\ 
$\bullet$ M09: Mermilliod, Mayor \& Udry~(\cite{mermilliod2009});  
          K14: Kovacs et al.~(\cite{kovacs2014});  
          R17: Rebull et al.~(\cite{rebull2017});  
          M15: Maxted et al.~(\cite{maxted2015}) + current literature (11 stars);  
          W17: Winn et al.~(\cite{winn2017});  
          A01: Abt~(\cite{abt2001});  
          Net17: Netopil et al.~(\cite{netopil2017}); 
          M13: Molenda-Zakowicz et al.~(\cite{molenda2013});  
          N15: Niemczura et al.~(\cite{niemczura2015});  
          N17: Niemczura et al.~(\cite{niemczura2017});  
          N13: Nielsen et al.~(\cite{nielsen2013});  
          R15: Reinhold \& Gizon~(\cite{reinhold2015});  
          Mc14: McQuillan et al.~(\cite{mcquillan2014}) 
\end{flushleft}
\end{table}

To estimate the inclination angle, we needed to derive stellar radii. To achieve 
this goal, we opted to use the IRFM-based $T_{\rm eff}$ calibration of 
Huang et al.~(\cite{huang2015}) from Johnson/2MASS $V-J$, $V-K$ colors. The 
observed magnitudes were dereddened by assuming $E(B-V)=0.027$ and standard 
extinction law (see Taylor~\cite{taylor2006} and Yuan et al.~\cite{yuan2013}). 
By considering the different reddenings used, our temperatures are consistent 
with those given by Cummings et al.~(\cite{cummings2017}) in their chemical 
abundance analysis of the same cluster.  These $T_{\rm eff}$ values were matched 
with the $Z=0.0200$, age$=0.8$~Gyr PARSEC isochrone of Marigo et al.~(\cite{marigo2017}). 
With $Z_{\odot}=0.0152$ used in the PARSEC models, the adopted metal abundance 
is consistent with those given in current spectroscopic studies (i.e., 
Boesgaard et al.~\cite{boesgaard2013};  see also 
Cummings et al.~\cite{cummings2017}). The age used is in agreement with the 
value currently predicted by models including the effect of stellar rotation 
(Brandt \& Huang~\cite{brandt2015}), but older than the standard age of 
$\sim 0.6$~Gyr (e.g., Fossati et al.~\cite{fossati2008}). Because  the stars 
studied in this work are far from the turn-off luminosity of the cluster, this 
difference has no effect whatsoever on the results presented in this paper. 
Once the observed and isochrone temperatures are matched, the radii are 
estimated from the model luminosities, using the blackbody formula.  

To check the consistency between the radii determined above and those 
derived by various other methods, first we took the database of 
Bourg\'es et al.~(\cite{bourges2014}, \cite{bourges2017}) containing 
photometrically calibrated stellar angular diameters from direct 
interferometric data. We found $74$ matches with our Praesepe sample. 
Assuming a cluster distance of $182$~pc 
(e.g., van~Leeuwen~\cite{vanleeuwen2009}), the angular diameter 
$\phi$ was converted to stellar diameter by using the formula 
$R/R_{\odot}=19.56\phi$~[mas]. These radii are plotted in Fig.~\ref{ldd} 
as a function of $T_{\rm eff}$. We see that the test dataset is 
rather sensitive to blending. Therefore, we think that the overall 
match of the isochrones with the part between the lower envelope and 
the ridge supports the radii values used in the present work, based 
on the [Fe/H]$=+0.12$ isochrone.  

In the second consistency check (Fig.~\ref{boyajian}) we performed 
a non-direct comparison by taking the interferometric data on $125$ 
Galactic field stars published by Boyajian et al.~(\cite{boyajian2013}). 
In addition to these data we added the planetary host samples of 
hot Jupiters (HJs) and Kepler planets (KOIs) from Table~\ref{datasets} 
and the compilation of $199$ planetary systems by 
Hartman et al.~(\cite{hartman2016}). Again, with the obviously older 
age and wide metallicity spread of these field star data, the isochrone 
used in this work for Praesepe (representing a relatively young and 
moderately metal-rich sample) seems to be well justified.

\subsection{Control samples}
We used five control samples to gain further support for the better assessment 
of the statistical significance of the rotation angle distribution derived 
for Praesepe. These samples (see Table~\ref{datasets}) were chosen 
under the assumption of near homogeneity of rotation axis distributions 
within each set. Because of the observed transits and the likely 
alignment of the stellar rotation and orbital angular momentum vectors, 
the two planet host samples are expected to show dominance of 
inclination angles close to $90^\circ$ (exceptions are the few 
non-aligned systems;  see Winn et al.~\cite{winn2017}). 
On the other hand, the magnetic, chemically peculiar (MCP) stars of 
Field~(A) and (B) are expected to have random rotational axis orientations, and 
therefore may serve as good control groups for testing the isotropic 
distribution. The same is true for the  spotted star sample Field~(C), 
but they are members of the Kepler field, and so there is basically no selection 
effect for the amplitude of the observable photometric variation. 
Therefore, objects with small inclination angles have an equal 
probability of being included in  the sample; however, the 
limited accuracy of ${{\rm v}sini}$ could still pose some constraints on slow rotators.       

Two selection criteria are applied to all items used in the analysis: 
i) avoiding poorly determined periods, stars with $P_{\rm rot} > 40$~days 
are omitted, and 
ii) avoiding clearly bad items, stars with $sini > 2.0$ are omitted.  
Except for Field (B), where the stellar radii are calculated from 
the tabulated luminosity and effective temperature, we used the 
radii given in the respective publications. For Field (B) there are $1347$ 
stars in the catalog of Netopil et al.~(\cite{netopil2017}), but only 
$193$ stars have all the necessary parameters for this study. The Field~(C) 
data resulted from the cross-correlation of several spectroscopic and 
photometric catalogs covering the Kepler field (see Table~\ref{datasets} 
for the respective publications). Radii for this dataset are as given 
in the corresponding catalogs or were  taken from Huber et al.~(\cite{huber2014}) 
when needed. 

%
%%%%%%%%%%%%%%%%%%%%%%%
% SECTION 3
%%%%%%%%%%%%%%%%%%%%%%%
%
\section{Method of analysis} 
In Sect.~4  we  first make a simple comparison of the Cumulative 
Distribution Functions CDFs for $sini$ derived from the observed ${{\rm v}sini}$, 
rotational period $P_{\rm rot}$, and estimated stellar radius $R_{\rm star}$. 
With the assumption of dominating equatorial spots in the derived  
rotational periods, we use the formula 
$sini={\rm v}sini\ P_{\rm rot}/(2\pi R_{\rm star})$ to estimate $sini$. 
Second, after a brief classification of the CDFs, following 
Jackson \& Jeffries~(\cite{jackson2010}), we model the observed 
distributions by assuming various distributions of $i$ and considering the 
distortion caused by the observational errors on these theoretical distributions. 
From the formula already quoted, we have 

%
%################
%  Eq. (1)
%################
%
\begin{eqnarray}
sini_{\rm obs} & = & {{{(1+\xi_{\rm P}/P_{\rm rot}^{\, 0})(1+\xi_{\rm {\rm v}sini}/{\rm v}sini^{\, 0})} \over {1+\xi_{\rm R}/R_{\rm star}^{\, 0}}}}sini 
\hskip 2mm . 
\end{eqnarray}
Here $P_{\rm rot}^{\, 0}$, $R_{\rm star}^{\, 0}$, and ${\rm v}sini^{\, 0}$ denote the 
{\em true} stellar parameters, which we  approximate with their observed/computed 
counterparts. The errors $\xi_{...}$ are Gaussian, except for 
$\xi_{\rm {\rm v}sini}$, which can be nicely represented by a LogNormal 
distribution. The noise parameters are taken from the observations (see below).  

In modeling the observed CDF with a given distribution of $sini$, we use 
Eq.~(1) with $50$ different realizations of 
$(sini,\xi_{\rm R},\xi_{\rm P},\xi_{{\rm v}sini})$ for all members of the 
sample. The CDFs of these $sini_{\rm obs}$ values are computed and 
an average CDF is obtained with the corresponding standard deviations 
derived from the $50$ realizations. The quality of the fit is parameterized 
by the RMS of the residuals between the target and the  average CDF. 

Choosing the size of the error components in the modeling 
is important in general, and could be crucial in the interpretation 
of the resulting underlying distribution of $sini$. This is especially 
true for $\xi_{{\rm v}sini}$. After an exhaustive series of tests, we 
decided to trust in the published errors and use them without any up- or 
down-scaling, depending on some consistency criteria that might be applied 
to the given dataset (e.g., using unreasonable down-scaling to reach 
better agreement with the expected aligned distribution for the 
planetary hosts). For the period and radius, the sources of the 
datasets include irregular noise information. Therefore, we decided 
to estimate the expected errors. Based on the comparison of the 
periods of Praesepe from Kovacs et al.~(\cite{kovacs2014}) and 
Rebull et al.~(\cite{rebull2017}), we found that the period errors are 
proportional to the period with a scaling factor of $0.03$. For the 
radius errors we use the publication of Winn et al.~(\cite{winn2017}). 
The radius errors given in this publication show a reasonably tight 
correlation with the radii. In summary, assuming standard Gaussian 
distributions, we use the following error formulae
%
%################
%  Eq. (2)
%################
%
\begin{eqnarray}
\xi_{\rm P} & = & {\rm GAUSS_{\rm P}}\times0.03 P_{\rm rot}^{\, 0} \hskip 2mm , \\ \nonumber
\xi_{\rm R} & = & {\rm GAUSS_{\rm R}}\times|0.27R_{\rm star}^{\, 0}-0.2| 
\hskip 2mm . 
\end{eqnarray}
We note that the results presented in this paper are only mildly sensitive 
to the assumed noise in the period and radius. The errors on ${\rm v}sini$ 
are far more important. 

Finally, the inclination angles are generated by following either the isotropic 
(Chandrasekhar \& M\"unch~\cite{chandra1950}, see also Cur\'e et al.~\cite{cure2014}) 
or a cone distribution (see Jackson \& Jeffries~\cite{jackson2010}). While 
the the isotropic model has no free parameters except  for the generally 
accepted noise model for the dataset, the cone distribution has two adjustable 
parameters: $\alpha$, the inclination of the axis of the cone, and $\lambda$, 
half of the opening angle of the cone. Best fit cone parameters are searched 
for by a simple grid search in $[0,\pi/2]$ for both parameters. The errors 
on these parameters come from the possible parameter regimes allowed by the 
$1\sigma$ scatter of the realization-dependent standard deviations 
of the residuals between the target CDF and the best fitting model. 
In the course of the error estimation, using the inverse of the CDF 
residuals as weights, the cone parameters are updated.

%
%%%%%%%%%%%%%%%%%%%%%%%
% SECTION 4
%%%%%%%%%%%%%%%%%%%%%%%
%
\section{Results}
\begin{figure}[!h]
 \vspace{0pt}
 \centering
 \includegraphics[angle=-90,width=85mm]{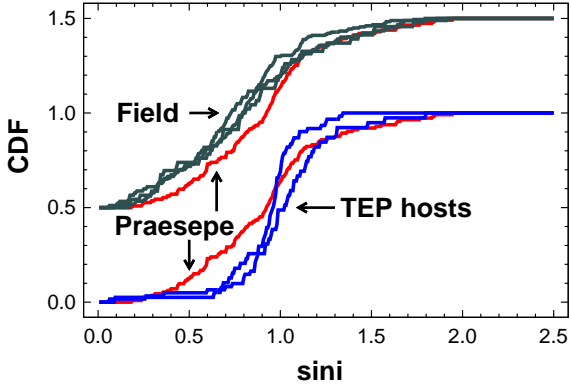}
 \caption{Comparison of the CDFs of {\it sini} obtained from the observed 
 ${\rm v}sini$, $P_{\rm rot}$, and estimated $R_{\rm star}$ values. For better 
 visibility, the CDFs are shifted vertically by $0.5$ with respect to those 
 of the planetary hosts (TEPs). The three gray lines correspond to the 
 three sets of Galactic field stars  given in Table~\ref{datasets}. 
 Similarly, the two blue lines are related to the exoplanet host stars. 
 The CDF for Praesepe is shown twice for easier comparison with the 
 CDFs corresponding to the different datasets.}
\label{obs_cdf}
\end{figure}

First we compare the distributions of the $sini$ values derived from the 
observations for the six datasets. Figure~\ref{obs_cdf} shows the 
result of this comparison, clearly indicating the expected similarity of 
the various groups of stars. To make this statement more quantitative, 
we perform a two-sample Kolmogorov--Smirnov (K-S) test for the pairs of 
the datasets. As expected, Table~\ref{K-S_contingency} confirms the 
classification suggested by Fig.~\ref{obs_cdf}. Although the relative 
errors of ${{\rm v}sini}$ vary by several factors over the samples, there 
are datasets of similar quality that show markedly different distributions, 
for example Praesepe and the planet hosts, and similar distributions with 
significantly different noise properties, for example  set (C) in the field 
star sample.\footnote{The relatively large K-S probability associated 
with the Praesepe - Field (C) samples is attributed in part to the 
small sample size of the latter.} Therefore, the differences 
in the distributions likely reflect real differences in the distributions 
of the inclination angles.  
 
%
%################
% Figure 3
%################
%

%
%================
% Table 2
%================
%
\begin{table}[!h]
%\centering
  \caption{Contingency table for the K-S statistics of the CDFs of 
   the observed $sini$}
  \label{K-S_contingency}
  \scalebox{0.79}{
  \begin{tabular}{lcccccc}
  \hline
 Dataset            & Praesepe    & HJ$_{\rm host}$ & KOI$_{\rm host}$ & Field~(A) & Field~(B) & Field~(C) \\
  \hline
Praesepe            & {\bf 1.000} & {\bf 0.103}     &      0.005  &      0.018  &      0.002  & {\bf 0.245}\\
HJ$_{\rm host}$     & {\bf 0.103} & {\bf 1.000}     &      0.028  &      0.000  &      0.000  &      0.005 \\
KOI$_{\rm host}$    &      0.005  &      0.028      & {\bf 1.000} &      0.000  &      0.000  &      0.000 \\
Field~(A)           &      0.018  &      0.000      &      0.000  & {\bf 1.000} & {\bf 0.374} & {\bf 0.620}\\
Field~(B)           &      0.002  &      0.000      &      0.000  & {\bf 0.374} & {\bf 1.000} & {\bf 0.410}\\
Field~(C)           & {\bf 0.245} &      0.005      &      0.000  & {\bf 0.620} & {\bf 0.410} & {\bf 1.000}\\
 \hline
\end{tabular}}
\begin{flushleft}
{\bf Notes:} 
The entries show the probabilities $P=Pr[d_{\rm max}>d_{\rm max}^{\rm obs}]$, 
where $d_{\rm max}^{\rm obs}$ is the observed maximum difference between the CDFs 
of the pair of datasets tested. Low probabilities indicate that the two CDFs are 
drawn from different distributions. For easier comparison, the $P>0.05$ cases are 
in bold. We use two-sample Kolmogorov--Smirnov statistics. 
\end{flushleft}
\end{table}

\begin{figure}
 \vspace{0pt}
 \centering
 \includegraphics[angle=0,width=85mm]{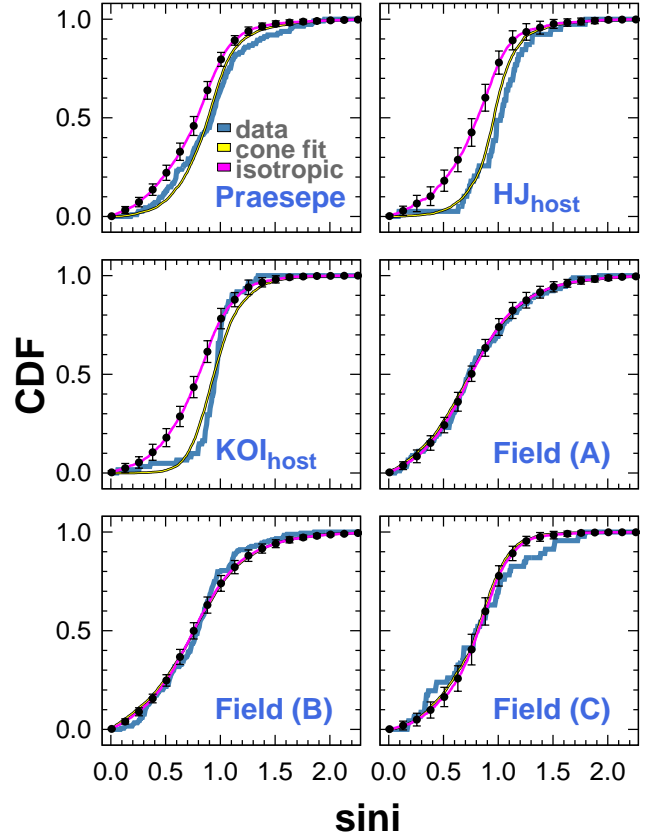}
 \caption{Observed cumulative distributions of $sini$ and the 
 associated model distributions assuming isotropic and cone 
 distributions (see Jackson \& Jeffries~\cite{jackson2010}) for 
 the inclination angle. See Table~\ref{cdf_match} for the 
 $(\alpha,\lambda)$ parameters used in the cone distributions.    
 The error bars correspond to $1\sigma$ limits derived from the 
 random simulations of the theoretical CDFs. To avoid crowding, 
 we show them only for the isotropic distributions because the errors 
  for the cone distributions are very similar.}
\label{cdf_all_fit}
\end{figure}

To model the underlying distribution of $sini$ we followed the methodology 
described in Sect.~3. Figure~\ref{cdf_all_fit} shows the result for the six 
datasets studied in this work. Starting with Praesepe, we see that the 
isotropic distribution is clearly distinct from the observed distribution.\footnote{In 
testing the robustness of this result against binaries and blends, we 
omitted the $28$ CMD outliers shown in Fig.~\ref{ldd} and found that the 
distinctiveness of the two distributions remains, albeit with a lower 
significance: the RMS between the two CDFs decreased from $0.076$ (full sample) 
to $0.062$ (without the CMD outliers). On the other hand, omitting only 
the $31$ spectroscopic binaries -- half of them are not outliers -- leads 
to an increase in the RMS to $0.095$.} However, the shape of the CDF is 
very similar to that of the observed CDF. This prompted us to investigate 
the possibility of the overestimation of the ${\rm v}sini$ values. 
The Mermilliod et al.~(\cite{mermilliod2009}) data come from the long-term 
monitoring of this cluster by the CORAVEL instrument. The correlation 
profiles were modeled by Gaussians, assuming broadening factors from rotation, 
stellar turbulence, and instrumental effects (see Benz \& Mayor~\cite{benz1984}). 
From the $\sigma$ of the fitted profile, ${\rm v}sini$ is computed from 
${\rm v}sini=A\sqrt{\sigma^2-\sigma_0^2}$. Here $A=1.9$ and $\sigma_0$ 
stands for the broadening of the non-rotating star. We find that a relatively 
small increase in $\sigma_0$ of $0.2$~kms$^{-1}$ makes the isotropic model 
valid. Although this is still considerably far from the expected range of 
$\sigma_0$ (see Fig.~1 in Queloz et al.~\cite{queloz1998}), it calls for 
 even higher precision data to give a more reliable answer 
on the issue of zero point of the $vsini$ data. On the other hand, by 
checking other (even though sparse and, in general, less accurate) data sources 
on Praesepe (Cummings et al.~\cite{cummings2017}, Quinn et al.~\cite{quinn2012}, 
Pace \& Pasquini~\cite{pace2004}, Boesgaard et al.~\cite{boesgaard2004}, 
Rachford~\cite{rachford1998}, Malavolta et al.~\cite{malavolta2016}, 
Mann et al.~\cite{mann2017}) we find that most  of the 
published values ($\sim 80$\%) are  an average of  $10$--$20$\%  
{higher} than our values. Therefore, we think that the current data lend 
support to a non-isotropic inclination angle distribution rather than 
to the isotropic one.    

For the other datasets we see a comfortable overall consistency 
between the expected distributions and the best fitting models. The 
optimal cone models (with large parameter scatter,  as expected) are 
indistinguishable from the isotropic model for all field star samples, 
although there is a slight indication for the underestimation of 
the observational noise for Field (C):  the observed CDF is shallower 
than that of the isotropic distribution. The opposite is likely to be 
true for the planetary host samples, especially for the KOI sample. 
We find that decreasing the noise on ${\rm v}sini$ by a factor of two 
(relative to the values given by Winn et al.~\cite{winn2017}), the fit 
improves substantially with ${\sigma_{\rm cone}=0.028\pm0.008}$ and 
cone parameters ${\alpha=80^{\circ}\pm11^{\circ}}$ and 
${\lambda=27^{\circ}\pm18^{\circ}}$. These cone parameters may imply 
the presence of some oblique systems in the KOI sample. Instead of 
making any far-reaching conclusion of this sort from this test alone, 
we refer to Hirano et al.~(\cite{hirano2014}) and Winn et al.~(\cite{winn2017}) 
for further discussion of the topic. However, we draw attention to the 
importance of proper noise estimation when modeling the distribution 
of the inclination angles from spectroscopic rotation velocities.  

%
%################
% Figure 4
%################
%

%
%================
% Table 3
%================
%
\begin{table}[!h]
%\centering
  \caption{Summary of the CDF model matches}
  \label{cdf_match}
  \scalebox{0.93}{
  \begin{tabular}{lcccc}
  \hline
 Dataset        & $\alpha$ & $\lambda$ & $\sigma_{\rm cone}$ & $\sigma_{\rm iso}$ \\
  \hline
Praesepe         & $76 \pm 14$ &   $47 \pm 24$ &  $0.033 \pm 0.011$ &  $0.079 \pm 0.017$ \\
HJ$_{\rm host}$  & $85 \pm 09$ &   $18 \pm 13$ &  $0.034 \pm 0.015$ &  $0.142 \pm 0.024$ \\
KOI$_{\rm host}$ & $81 \pm 14$ &   $24 \pm 23$ &  $0.049 \pm 0.011$ &  $0.128 \pm 0.018$ \\
Field~(A)        & $52 \pm 43$ &   $76 \pm 46$ &  $0.015 \pm 0.009$ &  $0.016 \pm 0.008$ \\
Field~(B)        & $55 \pm 46$ &   $69 \pm 58$ &  $0.025 \pm 0.007$ &  $0.031 \pm 0.009$ \\
Field~(C)        & $48 \pm 42$ &   $75 \pm 32$ &  $0.046 \pm 0.013$ &  $0.052 \pm 0.013$ \\
 \hline
\end{tabular}}
\begin{flushleft}
{\bf Notes:}\\
$\bullet$
The cone model parameters $\alpha$ and $\lambda$ denote the cone axis 
inclination angle and half of the opening angle, respectively. The standard 
deviations of the residuals between the model and the target CDFs are 
denoted by $\sigma_{\rm cone}$ and $\sigma_{\rm iso}$ for the cone and 
isotropic distribution, respectively. \\
$\bullet$
Errors have been computed from the Monte Carlo simulation, as described 
at the end of Sect.~3, and are subject to realization dependence at the 
$\sim 10$\% level.  
\end{flushleft}
\end{table}
% 

%%%%%%%%%%%%%%%%%%%%%%%
% Section 5
%%%%%%%%%%%%%%%%%%%%%%%
%
\section{Conclusions}
By using recently determined stellar rotation periods, available 
spectroscopic rotation velocities, and evolutionary stellar radii 
based on IRFM-calibrated effective temperatures, we derived 
the distribution of the line-of-sight inclination angles from 
$113$ main sequence F--K stars in the Praesepe open cluster. 
With the aid of statistical modeling, we found signatures of 
a broad rotational axis alignment. The cone model of 
Jackson \& Jeffries~(\cite{jackson2010}) yielded an overall cone 
axis angle of $76^\circ \pm 14^\circ$ and an opening angle 
of $47^\circ \pm 24^\circ$. Isotropic angle distribution 
is far less likely, except when the spectroscopic 
rotational velocities used in this work are systematically 
overestimated. Although we cannot exclude  this from being the 
case, a comparison made with other spectroscopic data available 
in the literature makes this possibility less likely. 

This is the first result on a possible alignment of stellar 
rotation axes on a large sample. In addition, the sample used in 
the present work is highly complete with regard to the available ${\rm v}sini$ 
data. Employing only about $40$ stars per cluster, earlier investigations 
by Jackson \& Jeffries~(\cite{jackson2010}) concluded with the 
low likelihood of alignment in the Pleiades and Alpha~Per clusters 
(see also the recent analysis by Jackson et al.~\cite{jackson2018}). 
To date, the only work suggesting stellar spin axis alignment in 
clusters is that of Corsaro et al.~(\cite{corsaro2017}). However, 
their result is based  on small samples (25 stars for NGC~6791 
and 23 stars for NGC~6819). 

The currently available data on Praesepe allow only a rather inaccurate 
estimation of the inclination angles. With the type of instrumentation used 
in the field of extrasolar planets, the same accuracy of $\sim 0.5$~kms$^{-1}$ 
in ${\rm v}sini$ could be reached. With a further improvement in the 
photometry and blend analysis, the error of the radius estimates could also 
be decreased  to $3-5$\%;  with the already accurate rotation 
periods, this  would enable us to estimate $sini$ with an accuracy of $\sim 7-10$\%. 
This is a two- to threefold decrease in error compared with what we have now.

Our findings in this work and those of Corsaro et al.~(\cite{corsaro2017}) are 
very difficult to understand by current stellar formation and cluster 
evolution theories. It is hard to see how some complicated magnetohydrodynamic 
effects can lead to even broadly organized spin axis distribution on a 
gigayear timescale after the star formation period. 

At this moment we think that the most effective way to make progress 
in the more secure disentangling of the various types of distribution is 
to acquire more accurate spectroscopic rotational velocities as described 
above. This, together with a better estimation of the noise budget, will 
most likely lead to  a cleaner observational 
input for the theory of rotational axis alignment.

%%%%%%%%%%%%%%%%%%%%%%%
% Acknowledgements
%%%%%%%%%%%%%%%%%%%%%%%
%
\begin{acknowledgements}
The thorough reports of the referee, leading to a deep revision of 
the methodology used and the conclusions drawn, are greatly appreciated.  
We would like to thank  Luisa Rebull for the prompt and instructive 
response to our inquiry about data availability before journal publication. 
Martin Netopil is acknowledged for his clarification on the error budget 
of the rotational velocities of mCP stars in their database. We thank  
Stephane Udry for his help in understanding the assignment of errors to 
the CORAVEL rotational velocities. This research has made use of the VizieR 
catalogue access tool, CDS, Strasbourg, France. 
\end{acknowledgements}

\bibliographystyle{aa} % style aa.bst

\end{document}